\documentclass[aps,prd,reprint,nofootinbib,superscriptaddress]{revtex4-2}
\usepackage{amsmath,amssymb,bm,graphicx}
\usepackage{caption}
\usepackage{subcaption}
\usepackage{color}
\usepackage[usenames,dvipsnames,svgnames]{xcolor}
\definecolor{mypurple}{rgb}{0.3,0,0.9}
\usepackage[colorlinks=true,linkcolor=blue,citecolor=blue,urlcolor=mypurple]{hyperref}
\newcommand{\be}{\begin{equation}}
\newcommand{\ee}{\end{equation}}
\newcommand{\ba}{\begin{eqnarray}}
\newcommand{\ea}{\end{eqnarray}}
\newcommand{\nn}{\nonumber\\}
\begin{document}

\title{Parton wakes and field angular momentum in chiral QCD plasma}

\author{Mohammad Yousuf Jamal}
\email{yousufjml5@gmail.com}
\affiliation{Institute of Particle Physics and Key Laboratory of Quark and
Lepton Physics (MOE), Central China Normal University, Wuhan 430079, China}

\begin{abstract}
A fast parton moving through the quark-gluon plasma trails a wake of color fields, usually characterized through its screening cloud and the associated energy loss. Its behavior in a plasma carrying a chiral imbalance, set by a chiral chemical potential $\mu_5$, remains largely unexplored, in particular whether the imbalance induces a parity-odd component in the wake. We address this within chiral kinetic theory with Berry curvature corrections, at finite $\mu_5$ and zero magnetic field. The screening sector is found to be nearly insensitive to $\mu_5$, which enters only through the Debye mass. The chiral imprint resides instead in a parity-odd sector, comprising an azimuthal field encircling the parton, a poloidal chromo-magnetic field, and a circulating color current. All three are linear in $\mu_5$ and transverse to the velocity, and hence do no work on the parton. As the central result, these handed fields endow the wake with a net field angular momentum about the parton direction, odd in $\mu_5$ and of order $10^{-4}$--$10^{-3}\,\hbar$ per parton. The chiral medium thereby opens a parity-odd channel connecting hard probes to the chiral and vortical structure of the plasma.
\end{abstract}
\maketitle

\section{Introduction.}  The quark-gluon plasma (QGP) is the deconfined state of strongly interacting matter produced in ultrarelativistic heavy-ion collisions at the Relativistic Heavy Ion Collider and the Large Hadron Collider. Its bulk evolution is well described by relativistic viscous hydrodynamics~\cite{gale_review,heinz_snellings,romatschke}, and the transport coefficients that enter this description, such as the shear and bulk viscosities, are now extracted from flow data through Bayesian methods~\cite{bernhard,jetscape}, as the field enters a precision era. A complementary picture of the medium comes from hard probes, the energetic partons such as jets and heavy quarks that are produced in the early stages of the collision and lose energy through collisional and radiative processes as they traverse the plasma~\cite{mustafa2005,qin_prl}. Heavy quarks serve as particularly clean probes, since their large mass tags them throughout the evolution~\cite{cao,sebastian_jamal}.

A related observable is the wake that the parton induces in the medium. A relativistic color charge sets up a disturbance of color charge density, color current, and induced chromo-electromagnetic fields~\cite{ruppert_muller,chakraborty_prd,chakraborty_prc}. Within linear response and the hard-thermal-loop approximation, the parton follows Wong's equations, and the induced field is determined by the medium gluon self-energy~\cite{thoma_gyulassy,wong}. A slow parton drags a comoving screening cloud, while a fast one generates a Mach-cone structure behind it. The wake thus reflects the collective response of the medium, and it has been studied in isotropic, anisotropic, collisional, and nonequilibrium plasmas~\cite{dumitru,carrington2015,jamal_das_ruggieri,romatschke_strickland}.

In parallel, the chiral anomaly has drawn wide interest in QGP studies~\cite{kharzeev_review}. Topological fluctuations produce a local imbalance between right- and left-handed quarks, set by a chiral chemical potential $\mu_5\equiv\mu_R-\mu_L$. In a magnetic field, this imbalance drives the chiral magnetic effect, a current along the field~\cite{fukushima_kharzeev_warringa,kharzeev_son}, searched for through charge-separation observables~\cite{star_cme,alice_cme}. The macroscopic dynamics of chiral fermions is captured by chiral kinetic theory with Berry curvature corrections~\cite{stephanov_yin,son_yamamoto,chen_ishii}. The imbalance also modifies the soft sector itself, even in the absence of a magnetic field. The gluon self-energy acquires a parity-odd, antisymmetric part linear in $\mu_5$, which splits the transverse modes into right- and left-handed circular branches, drives the chiral plasma instability, and strengthens the Debye screening~\cite{akamatsu_yamamoto,carignano_manuel,ghosh_jamal_kurian}. A parton traversing such a medium polarizes it and trails a wake of color fields, whose chiral, parity-odd part and the angular momentum it carries are the subject of this work.

Interest in such handed effects is reinforced by the strong vorticity of the plasma. In non-central collisions, the colliding nuclei carry a large orbital angular momentum, of order $10^{5}\,\hbar$, a sizable fraction of which is retained by the medium rather than carried away by the spectators. Part of this angular momentum manifests as a global spin polarization of emitted $\Lambda$ hyperons, measured by STAR~\cite{star_polarization,liang_wang,becattini_lisa}. The polarization follows the thermal vorticity of the flow, identifying the QGP as the most vortical fluid known and rendering the storage and transfer of angular momentum a central open question. Vorticity and chirality are further linked through anomalous transport, such as the chiral vortical effect, in which vorticity drives a current along the rotation axis~\cite{kharzeev_review}. This picture concerns primarily the bulk medium; how a hard parton couples to the angular momentum and the chirality of the plasma is far less explored. The wake of such a parton provides a direct probe, since in a chiral medium it is itself handed.

In this work, we compute the wake of a fast parton in a chirally imbalanced QGP, at finite $\mu_5$ and zero magnetic field. We identify where the chiral imbalance leaves its mark on the wake and show that it resides in a distinct parity-odd sector rather than in the screening cloud, the latter being nearly independent of $\mu_5$ and sensitive to it only through the Debye mass. Natural units $\hbar=c=k_B=1$ are used throughout.

\section{Framework}

\subsection{Chiral response.} We treat the energetic parton as a classical particle
carrying an $SU(N_c)$ color charge, moving with velocity $\mathbf v$ on a
straight line through the medium. Its motion is governed by Wong's
equations~\cite{wong}, and the disturbance it leaves in the medium follows from
the linearized Yang-Mills equations~\cite{thoma_gyulassy}. As it moves, the
parton produces a color current,
\be
j^{j}_{a}(K)=2\pi g\,q_a\,v^{j}\,\delta(\omega-\mathbf k\cdot\mathbf v),
\ee
where $K=(\omega,\mathbf k)$, $q_a$ is the parton color charge, $g$ is the strong coupling 
and the delta function ties the response to the trajectory $\omega=\mathbf k\cdot\mathbf v$.
This current polarizes the medium, and the induced chromo-electric field is set by the effective gluon propagator
$\Delta^{ij}$ as
\be
E^{i}_{a}(K)=i\omega\,\Delta^{ij}(K)\,j^{j}_{a}(K),
\label{eq:Efield}
\ee
which carries the medium response through $\Delta$. The induced field is set by the effective gluon propagator $\Delta^{ij}$, obtained from the Dyson-Schwinger equation~\cite{thoma_gyulassy,ghosh_jamal_kurian},
\be
(\Delta^{-1})^{ij}=(k^2-\omega^2)\delta^{ij}-k^i k^j+\Pi^{ij},
\label{eq:dyson}
\ee
where $\Pi^{ij}$ is the spatial gluon self-energy. At finite $\mu_5$ and zero magnetic field,
\be
\Pi^{ij}=\Pi_T A^{ij}+\Pi_L B^{ij}+\Pi_A C^{ij},
\label{eq:Pi}
\ee
with the transverse, longitudinal, and antisymmetric projectors $A^{ij}=\delta^{ij}-\hat k^i\hat k^j$, $B^{ij}=\hat k^i\hat k^j$, and $C^{ij}=i\epsilon^{ijl}\hat k^l$.
The parity-even form factors are the hard-thermal-loop
results~\cite{thoma_gyulassy,bellac},
\ba
\Pi_L&=&-m_D^2\,\frac{\omega^2}{k^2}\Big[1-\frac{\omega}{2k}\mathcal L\Big],
\label{eq:PiL}\\
\Pi_T&=&\;\;m_D^2\,\frac{\omega^2}{2k^2}\Big[1-\frac{\omega^2-k^2}{2\omega k}
\mathcal L\Big],
\label{eq:PiT}
\ea
with $\mathcal L=\ln[(\omega+k)/(\omega-k)]$ and the retarded prescription
$\omega\to\omega+i0^+$. The parity-odd form factor is linear in $\mu_5$ and comes
from the Berry-curvature term in the chiral kinetic
theory~\cite{akamatsu_yamamoto,ghosh_jamal_kurian},
\be
\Pi_A\equiv C_A=\frac{\mu_5 g^2 k}{4\pi^2}\Big[1-\frac{\omega^2}{k^2}
\Big(1-\frac{\omega}{2k}\mathcal L\Big)\Big].
\label{eq:CA}
\ee
The chiral imbalance also raises the Debye mass~\cite{ghosh_jamal_kurian},
\be
m_D^2=(N_f+2N_c)\,\frac{g^2T^2}{6}+N_f\,\frac{g^2}{2\pi^2}\,\frac{\mu_5^2}{2},
\label{eq:mD}
\ee
so the screening length $1/m_D$ decreases as $\mu_5$ grows (see Appendix \ref{sec:figs} ). The antisymmetric
tensor $C^{ij}$ introduced in Eq.~(\ref{eq:Pi}) rotates a transverse vector by
ninety degrees about $\hat{\mathbf k}$, and squares to the transverse projector,
$C^2=A$. This property makes Eq.~(\ref{eq:dyson}) invertible in closed form, giving
\be
\Delta^{ij}=\frac{B^{ij}}{C_L}+\frac{C_T A^{ij}-C_A C^{ij}}{C_T^2-C_A^2},
\label{eq:Delta}
\ee
with $C_L=-\omega^2+\Pi_L$ and $C_T=k^2-\omega^2+\Pi_T$. On the
circular-polarization basis, the transverse block is diagonal, see Appendix \ref{sec:prop}, $\Delta_\pm=1/(C_T
\mp C_A)$, the circular birefringence of the chiral
medium~\cite{akamatsu_yamamoto,carignano_manuel}.

\subsection{Wake fields.}  Contracting the propagator of Eq.~(\ref{eq:Delta}) with $\mathbf v$ in
Eq.~(\ref{eq:Efield}) on the wake shell $\omega=\mathbf k\cdot\mathbf v$, splits the induced field into three pieces,
\ba
\mathbf E&=&\mathbf E_L+\mathbf E_T+\mathbf E_\chi,
\label{eq:Esplit}\\
\mathbf E_L&=&iQ\,\frac{\omega^2}{k\,C_L}\,\hat{\mathbf k},
\label{eq:EL}\\
\mathbf E_T&=&iQ\,\omega\,\frac{C_T}{C_T^2-C_A^2}\,\mathbf v_\perp,
\label{eq:ET}\\
\mathbf E_\chi&=&Q\,\omega\,\frac{C_A}{C_T^2-C_A^2}\,
\frac{\mathbf v\times\mathbf k}{k},
\label{eq:Echi}
\ea
where $\mathbf v_\perp=\mathbf v-\hat{\mathbf k}(\hat{\mathbf k}\cdot\mathbf v)$
and $Q\equiv g q_a$. The longitudinal field $\mathbf E_L$ is the screening field.
It alone sources the induced charge density through the wake potential
$\Phi(\mathbf k)=Q/(k^2\varepsilon_L)$ with $\varepsilon_L=-C_L/\omega^2$, so
$\varrho_{\rm ind}=k^2\Phi$ depends on $\mu_5$ only through $m_D$ inside $C_L$.
The transverse field $\mathbf E_T$ is the collective field of the medium. 
It is parity even, depending on $\mu_5$ only through $C_A^2$, and is unchanged
in character by the chiral imbalance, so we do not consider it further. The chiral field $\mathbf
E_\chi$ is the new piece, and it carries three exact properties (for details see Appendix~\ref{sec:decomp}).
It sources no charge, since $\varrho_{\rm ind}=i\,\mathbf k\cdot\mathbf E_\chi\propto\mathbf k\cdot(\mathbf v\times\mathbf k)=0$.
 It is
solenoidal and reverses under $\mu_5\to-\mu_5$, since $C_A$ is odd while $C_L$,
$C_T$, and $C_A^2$ are even. And it does no work on the parton, since $\mathbf v
\cdot(\mathbf v\times\mathbf k)=0$. Being azimuthal, $\mathbf E_\chi$ generates a poloidal chromo-magnetic field $\mathbf B_\chi$, lying in the meridional $(\rho,z')$ plane with no azimuthal component shown in Appendix~\ref{sec:bj},
\be
\mathbf B_\chi=Q\,\frac{C_A}{C_T^2-C_A^2}\,\frac{1}{k}
\big(k^2\mathbf v-\omega\,\mathbf k\big),
\label{eq:Bchi}
\ee
odd in $\mu_5$ and absent without chirality. The same response drives an
azimuthal color current; from $\mathbf j=i(\mathbf k\times\mathbf B+\omega\mathbf
E)$ only the chiral part has an azimuthal projection, so
\be
j_\phi\propto\frac{C_A}{C_T^2-C_A^2}\,(\omega^2-k^2),
\label{eq:jphi}
\ee
odd in $\mu_5$ and vanishing at $\mu_5=0$. It is the wake-driven analog of a
chiral magnetic current, generated by the parton itself rather than by an
external field.
\section{Results}
\subsection{The wake in position space.} We transform the induced fields to position
space using cylindrical coordinates comoving with the parton, with $z'=z-vt$ along
the direction of motion and $\rho$ the perpendicular distance from the axis. In
this frame the parton sits at the origin and the wake is stationary. With
$\mathbf v=v\hat{\mathbf z}$ and $\omega=vk_z$, the azimuthal angle integrates out
through $\int d\psi\,e^{ik_\perp\rho\cos\psi}=2\pi J_0(k_\perp\rho)$ and
$\int d\psi\,e^{ik_\perp\rho\cos\psi}\cos\psi=2\pi i J_1(k_\perp\rho)$, leaving a
Hankel transform in $k_\perp$ and a Fourier transform in $k_z$ (for details see Appendix \ref{sec:coord}). The induced color
charge density follows from the longitudinal field alone,
\be
\varrho_{\rm ind}=\frac{Q}{(2\pi)^2}\,\mathrm{Re}\!\int\! dk_\perp\,k_\perp
J_0(k_\perp\rho)\!\int\! dk_z\,e^{ik_z z'}\Big[-\frac{\omega^2}{C_L}\Big],
\label{eq:rho_pos}
\ee
which depends on $\mu_5$ only through $C_L$, hence only through $m_D$. The
azimuthal color current retains only the chiral projection of the response,
\ba
j_\phi=&-\frac{Qv}{(2\pi)^2}\,\mathrm{Re}\!\int\! dk_\perp\,k_\perp^2
J_1(k_\perp\rho)\nn
&\times \int\! dk_z\,e^{ik_z z'}\,\frac{\omega^2-k^2}{k}\,
\frac{C_A}{C_T^2-C_A^2}.
\label{eq:jphi_pos}
\ea

\begin{figure}[t]
  \centering
  \includegraphics[width=\columnwidth,height=0.50\columnwidth]{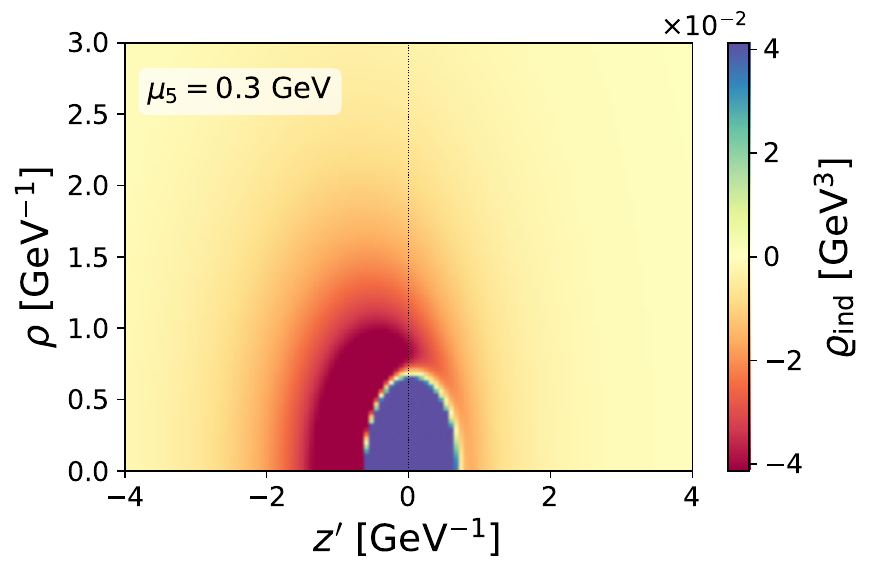}
  \caption{Induced color charge density $\varrho_{\rm ind}$ at $\mu_5=0.3$~GeV in the comoving plane; $z'$ along the parton motion, $\rho$ the transverse distance. The parton moves toward $+z'$.}
\label{fig:charge}
\end{figure}

\begin{figure}[t]
  \centering
  \includegraphics[width=\columnwidth,height=0.50\columnwidth]{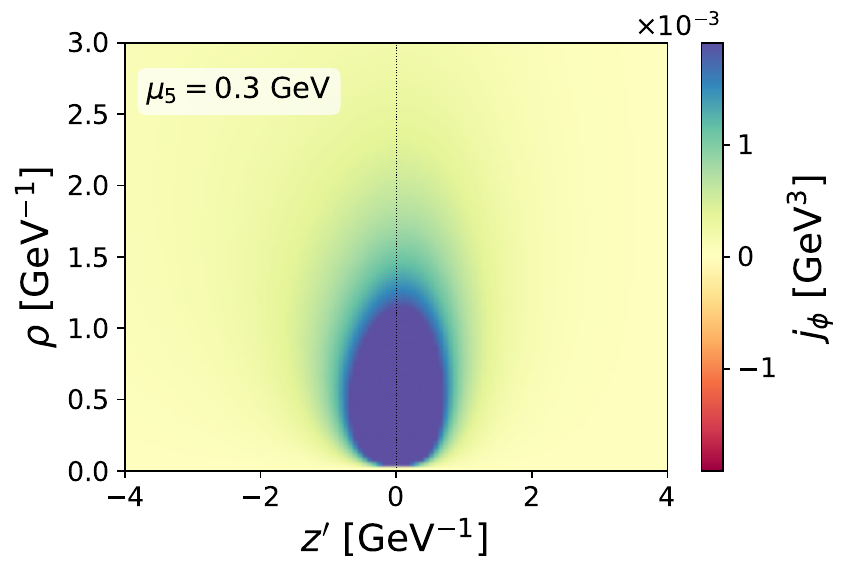}
  \caption{Chiral azimuthal color current $j_\phi$ at $\mu_5=0.3$~GeV. Coordinates as in Fig.~\ref{fig:charge}.}  \label{fig:jphi}
\end{figure}

To plot the results, we take $T=0.3$~GeV, $\alpha_s=0.3$, and $v=0.55$ throughout.
In the two-dimensional plots, the integrand is multiplied by a Gaussian factor
$e^{-(k/\Lambda)^2}$ with $\Lambda=7$~GeV, which suppresses the highest momenta
and removes grid artifacts; it is applied only to these plots, not to the
quantitative results. Fig.~\ref{fig:charge} shows the induced color charge
density. A core of one sign sits right at the parton, surrounded by the opposite
sign. This is the medium rearranging its color charge to screen the moving parton.
The screening charge trails the parton, leaning behind it toward negative $z'$.
Fig.~\ref{fig:jphi} shows the azimuthal color current $j_\phi$. It circulates
about the parton axis with a single sense of rotation, concentrated in the same
region as the screening charge, and is the chiral, parity-odd component of the
wake. It is present only for a nonzero chiral imbalance, and reverses when $\mu_5$
changes sign. Increasing the imbalance from $\mu_5=0.3$ to $0.5$~GeV leaves the
charge density essentially unchanged, below $0.1\%$, since it enters only through
$m_D$. The current instead scales almost linearly with $\mu_5$, growing by about
$67\%$ from $\mu_5=0.3$ to $0.5$~GeV, as expected for a quantity proportional to
$\mu_5$.

\begin{figure}[t]
  \centering
  \includegraphics[width=\columnwidth,height=0.50\columnwidth]{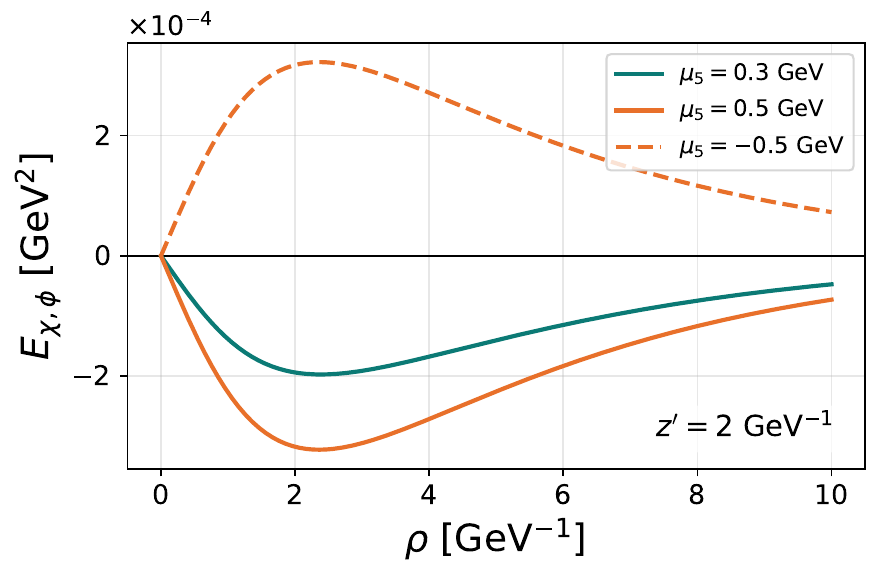}
 \caption{Azimuthal chiral field $E_{\chi,\phi}$ at $z'=2$~GeV$^{-1}$ for $\mu_5=0.3$, $0.5$, and $-0.5$~GeV (dashed). Parameters as in Fig.~\ref{fig:charge}.}
\label{fig:chiralfield}
 \end{figure}

\begin{figure}[t]
  \centering
  \includegraphics[width=\columnwidth,height=0.50\columnwidth]{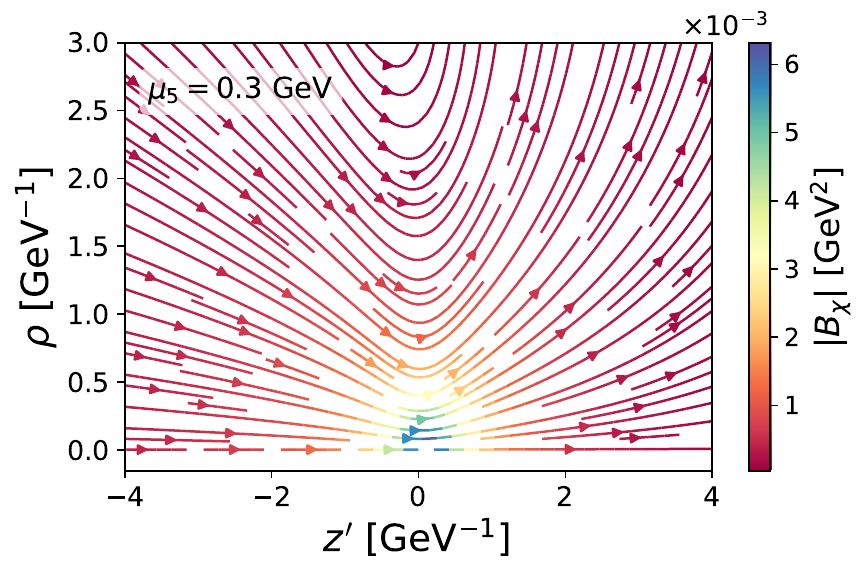}
  \caption{Chiral magnetic field $\mathbf B_\chi$, Eq.~(\ref{eq:Bchi}), in the meridional plane at $\mu_5=0.3$~GeV; streamlines show $(B_\rho,B_z)$ and color the magnitude $|\mathbf B_\chi|$. Parameters as in Fig.~\ref{fig:charge}.}
\label{fig:bpol}
\end{figure}

\subsection{The chiral field.} Fig.~\ref{fig:chiralfield} shows the chiral field
$E_{\chi,\phi}$ of Eq.~(\ref{eq:Echi}) at fixed $z'=2$~GeV$^{-1}$. It vanishes on
the axis, rises to a maximum near $\rho\simeq2$~GeV$^{-1}$, and decays at larger
$\rho$. On a common normalization, the chiral field is small, of order $10^{-3}$ of
the screening field, but it is the entire parity-odd content of the wake. It grows
with $\mu_5$, as seen from the $\mu_5=0.3$ and $0.5$~GeV curves, and reverses sign
for $\mu_5<0$, shown by the dashed curve. This reversal is the parity-odd
signature of the chiral wake, and it follows directly from $E_{\chi,\phi}\propto
C_A\propto\mu_5$. We found that the field is linear in $\mu_5$ over the
range studied and that it keeps the same shape at other $z'$ slices. 
The chiral field also generates the poloidal chromo-magnetic field $\mathbf B_\chi$, Eq.~(\ref{eq:Bchi}), shown in Fig.~\ref{fig:bpol} at $\mu_5=0.3$~GeV. The field lies in the $(\rho,z')$ plane, is strongest on the axis near $z'\simeq0$, and decays outward, with $B_\rho$ reversing across $z'=0$. Its cylindrical components $B_z,B_\rho$ are given below in Eqs.~(\ref{eq:Bz})--(\ref{eq:Brho}). It is the magnetic partner of the azimuthal chiral field, is linear in $\mu_5$, and reverses with the sign of $\mu_5$.

\subsection{Field angular momentum.} 
\begin{figure}[t]
  \centering
  \begin{subfigure}{0.49\columnwidth}
    \includegraphics[width=\textwidth,height=0.86\textwidth]{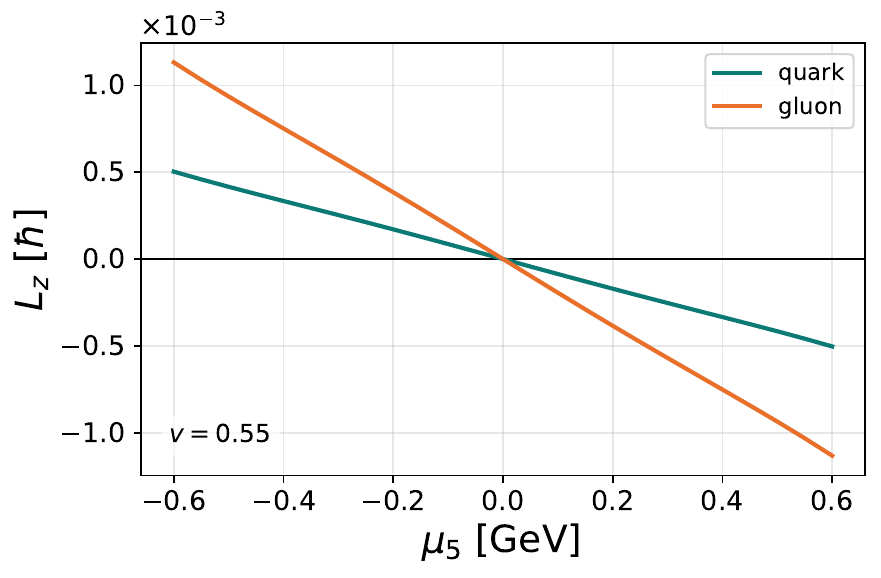}
    \caption{}\label{fig:lzmu}
  \end{subfigure}\hfill
  \begin{subfigure}{0.49\columnwidth}
    \includegraphics[width=\textwidth,height=0.86\textwidth]{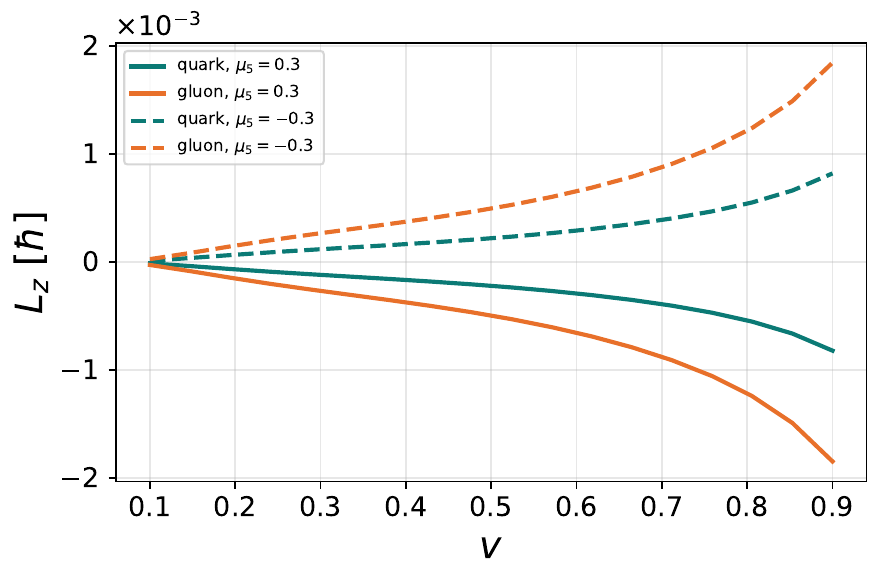}
    \caption{}\label{fig:lzv}
  \end{subfigure}
  \caption{Field angular momentum $L_z$ in units of $\hbar$, for a quark and a gluon. (a) Versus $\mu_5$ at fixed $v=0.55$. (b) Versus parton speed for $\mu_5=0.3$ and $0.5$~GeV. The soft cutoff is $k_{\max}=m_D(\mu_5,T)$. Parameters as in Fig.~\ref{fig:charge}.}
\label{fig:Lz}
\end{figure}

The handed configuration stores angular
momentum about the parton direction. In the same cylindrical reduction, with
$d_z=vk_z^2/(k^2C_L)+(C_T/(C_T^2-C_A^2))vk_\perp^2/k^2$ and $g_E=1/(k^2C_L)-C_T/
(k^2(C_T^2-C_A^2))$, the meridional electric and poloidal magnetic components are
\ba
E_z&=&\frac{Q}{(2\pi)^2}\mathrm{Re}\!\int\! dk_\perp k_\perp J_0\!\int\! dk_z
e^{ik_z z'}\,(i\omega d_z),
\label{eq:Ez}\\
E_\rho&=&-\frac{Qv^2}{(2\pi)^2}\mathrm{Re}\!\int\! dk_\perp k_\perp^2 J_1\!\int\!
dk_z e^{ik_z z'}\,k_z^2 g_E,
\label{eq:Erho}\\
B_z&=&\frac{Qv}{(2\pi)^2}\mathrm{Re}\!\int\! dk_\perp k_\perp J_0\!\int\! dk_z
e^{ik_z z'}\,\frac{k_\perp^2}{k}\frac{C_A}{C_T^2-C_A^2},
\label{eq:Bz}\\
B_\rho&=&-\frac{Qv}{(2\pi)^2}\mathrm{Re}\!\int\! dk_\perp k_\perp iJ_1\!\int\!
dk_z e^{ik_z z'}\,\frac{k_z k_\perp}{k}\frac{C_A}{C_T^2-C_A^2}.\nn
\label{eq:Brho}
\ea
The angular momentum stored in the wake field about the parton direction is
\be
L_z=\int d^3r\,[\mathbf r\times(\mathbf E\times\mathbf B)]_z
=2\pi\!\int\! dz'\!\int\!\rho^2 d\rho\,(E_z B_\rho-E_\rho B_z),
\label{eq:Lz}
\ee
with $\mathbf E\times\mathbf B$ the field momentum density. \footnote{The spatial structure of the integrand $\rho^2(E_z B_\rho-E_\rho B_z)$ is shown in the appendix ~\ref{sec:coord}, where the $\rho^2$ weight suppresses the contribution on the axis and concentrates it in a shell at intermediate $\rho$.} The meridional electric field is even in $\mu_5$ while the poloidal magnetic field is odd, so the integrand is odd in $\mu_5$, and $L_z$ vanishes identically at $\mu_5=0$. A nonzero $L_z$ is therefore a direct, parity-odd measure of the chiral imbalance, generated entirely by it and fixed in sign by the sign of $\mu_5$. Each field is linear in the source $Q=g q_a$, so $L_z\propto Q^2=g^2 q_a^2$, and the sum over color charges supplies the quadratic Casimir $C_2(R)$, larger for a gluon than for a quark. The momentum integral extends to the soft scale $k_{\max}=m_D(\mu_5,T)$, the boundary of the soft, classical-field regime, evaluated at the temperature and imbalance of each calculation. For $\alpha_s=0.3$, $T=0.3$~GeV, $v=0.55$, and $\mu_5=0.3$~GeV,
$L_z\simeq-2.5\times10^{-4}\,\hbar\ (\text{quark}),\quad -5.7\times10^{-4}\,\hbar\ (\text{gluon}),$ rising in magnitude to $-4.2\times10^{-4}\,\hbar\ (\text{quark})$ and $-9.4\times10^{-4}\,\hbar\ (\text{gluon})$ at $\mu_5=0.5$~GeV; the values are stable under refinement of the numerical grid. As shown in Fig.~\ref{fig:Lz}, $L_z$ is odd in $\mu_5$, vanishes at $\mu_5=0$, is nearly linear over the range studied, and grows monotonically with the parton speed, reflecting the stronger dynamical polarization of a faster parton. The calculation remains below the onset of the chiral plasma instability~\cite{akamatsu_yamamoto}, monitored throughout discussed in Appendix~\ref{sec:prop}.

The physical content of this result is that a hard parton crossing a chiral domain spins up the color field around it, depositing a definite, sign-selected angular momentum into the medium. This is the same quantity that the spin-polarization program accesses through the thermal vorticity of the bulk~\cite{star_polarization,becattini_lisa}, but here it is sourced locally by the parton and selected by the chirality of the medium rather than by the global orbital angular momentum of the collision. Because the magnitude and sign of $L_z$ track $\mu_5$ along the parton trajectory, the handed wake offers a parton-level route to local polarization, complementary to the bulk vorticity and carried by the hard probes already measured at the Relativistic Heavy Ion Collider and the Large Hadron Collider.

\section{Summary and outlook.} We have computed the wake of a fast parton in a chirally imbalanced quark-gluon plasma, at finite $\mu_5$ and zero magnetic field, within chiral kinetic theory with Berry-curvature corrections. The screening sector is found to be nearly insensitive to the imbalance, which enters only through the Debye mass. The chiral imprint resides instead in a parity-odd sector, an azimuthal chromo-electric field, a poloidal chromo-magnetic field, and a circulating color current, all linear in $\mu_5$, transverse to the velocity, and doing no work on the parton. As the central result, these handed fields endow the wake with a net field angular momentum about the parton direction, odd in $\mu_5$, of order $10^{-4}$--$10^{-3}\,\hbar$ per parton, and fixed in sign by the local chiral charge. To the best of our knowledge, this is the first identification of a field angular momentum carried by a parton wake, and the first demonstration that a hard probe couples to the chiral structure of the medium through a parity-odd, energy-conserving channel.

This result reframes how a hard parton connects to the rotational structure of the plasma. The spin-polarization program reads the angular momentum of the medium through the bulk thermal vorticity inherited from the global orbital angular momentum of the collision~\cite{star_polarization,becattini_lisa}. The mechanism found here is distinct and local, a parton deposits angular momentum directly into the surrounding color field as it crosses a chiral domain, providing a parton-level, parity-odd route to the same observable. It is, in effect, a microscopic and chirality-selective counterpart of the bulk vorticity that drives global polarization~\cite{liang_wang,star_polarization,becattini_lisa,alice_polarization}, accessible through the hard probes already measured at the Relativistic Heavy Ion Collider and the Large Hadron Collider.

The present result invites several extensions, which we intend to pursue. First, the computed $L_z$ is the angular momentum stored in the chromo-fields; translating it into the spin polarization of emitted hadrons requires coupling the wake to a transport and hadronization stage, in which the field angular momentum is transferred to medium quarks and then to final-state baryons. This connects the present mechanism to the vortical dynamics of the bulk medium, where the local angular momentum and its conversion to hadron spin have been studied extensively~\cite{becattini_lisa,captain_vorticity}. Second, the magnitude found here, of a few times $10^{-4}\,\hbar$ per parton, is modest relative to a hadron spin, so a quantitative signal will accumulate from the many hard partons traversing a chiral domain rather than from a single one; estimating this collective contribution is a natural next step. We further plan to extend the calculation to the combined chiral and magnetic case~\cite{ghosh_jamal_kurian,miransky_shovkovy}, to a collisional and anisotropic medium~\cite{carrington2017,romatschke_strickland}, and to a dynamically evolving $\mu_5$~\cite{kharzeev_review}. The parton wake thus emerges as a new and direct probe of the interplay between chirality, vorticity, and hard probes in the quark-gluon plasma.
\\
\section*{Acknowledgment.} The author acknowledges Central China Normal University (CCNU), Wuhan, China for the postdoctoral fellowship that supported this work.

\appendix

\section{Screening Mass}\label{sec:figs}

\begin{figure}[t]
  \centering
  \includegraphics[width=0.82\columnwidth,height=0.50\columnwidth]{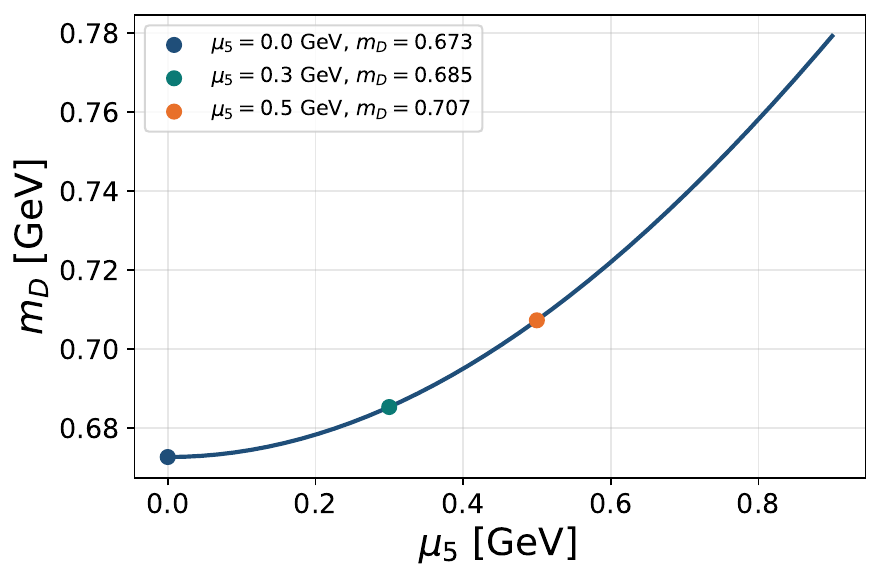}
  \caption{Chiral Debye mass $m_D$ as a function of $\mu_5$ from Eq.~(8) of the
  main text, for $T=0.3$~GeV, $\alpha_s=0.3$, $N_c=3$, $N_f=2$. Markers indicate
  $\mu_5=0$, $0.3$, and $0.5$~GeV. The mass sets the screening length $1/m_D$ and
  the soft cutoff $k_{\max}=m_D$ used in the angular momentum.}
  \label{fig:debye}
\end{figure}

The chiral Debye mass, Eq.\eqref{eq:mD} of the main text, is shown in Fig.~\ref{fig:debye};
it grows slowly with $\mu_5$ and sets both the screening length $1/m_D$ and the
soft cutoff $k_{\max}=m_D$. The chirality independence of the screening sector is
quantified in Fig.~\ref{fig:pfrac}, the induced potential changes by less than one
percent over the range of $\mu_5$ studied, confirming that the screening response
is sensitive to $\mu_5$ only through $m_D$.

\section{Propagator inversion and helicity diagonalization}
\label{sec:prop}

The projectors $A^{ij}=\delta^{ij}-\hat k^i\hat k^j$, $B^{ij}=\hat k^i\hat k^j$,
and $C^{ij}=i\epsilon^{ijl}\hat k^l$ form a closed algebra. With $\delta^{ij}=
A^{ij}+B^{ij}$ and $k^i k^j=k^2 B^{ij}$ one has
\ba
A^2&=&A,\quad B^2=B,\quad AB=BA=0,\nn && AC=CA=C,\quad BC=CB=0,
\ea
and the central identity
\ba
(C^2)^{il}&=&(i\epsilon^{ijm}\hat k^m)(i\epsilon^{jln}\hat k^n)\nn
&=&-(\delta^{ml}\delta^{in}-\delta^{mn}\delta^{il})\hat k^m\hat k^n
=\delta^{il}-\hat k^i\hat k^l=A^{il},\nn
\ea
so that $C$ acts as a ninety-degree rotation within the transverse plane and
squares to the transverse projector. Using the form factors of the main text, the
inverse propagator of Eq.\eqref{eq:Delta} becomes
\be
(\Delta^{-1})^{ij}=C_L B^{ij}+C_T A^{ij}+C_A C^{ij},
\ee
with $C_L=-\omega^2+\Pi_L$, $C_T=k^2-\omega^2+\Pi_T$, and $C_A=\Pi_A$. The
longitudinal sector decouples because $BA=BC=0$, giving $\Delta_L=B/C_L$. On the
transverse subspace we set $\Delta_\perp=\alpha A+\beta C$ and impose
$\Delta^{-1}\Delta_\perp=A$,
\ba
(C_T A+C_A C)(\alpha A+\beta C)&=&(C_T\alpha+C_A\beta)A+(C_T\beta\nn &&+C_A\alpha)C
=A,
\ea
where $C^2=A$ has been used. This yields $\alpha=C_T/(C_T^2-C_A^2)$ and
$\beta=-C_A/(C_T^2-C_A^2)$, so that
\be
\Delta^{ij}=\frac{B^{ij}}{C_L}+\frac{C_T A^{ij}-C_A C^{ij}}{C_T^2-C_A^2},
\ee
which is Eq.~\eqref{eq:Delta} of the main text. On the circular-polarization basis
$\hat{\mathbf e}_\pm=(\hat{\mathbf e}_1\pm i\hat{\mathbf e}_2)/\sqrt2$, with
$C\hat{\mathbf e}_\pm=\mp\hat{\mathbf e}_\pm$, the transverse block is diagonal,
\be
\Delta_\pm=\frac{1}{C_T\mp C_A},
\ee
the circular birefringence of the chiral medium. The two transverse modes are
degenerate at $\mu_5=0$ and are split by $C_A$ into right- and left-handed
branches; the longitudinal mode has no degenerate partner, so the splitting is
purely transverse. The conditions $C_T\mp C_A=0$ mark the onset of the chiral
plasma instability~\cite{akamatsu_yamamoto,carignano_manuel}. We verify $\min_k\mathrm{Re}(C_T^2-C_A^2)>0$ throughout, so the medium stays below this instability.
 
\section{Field decomposition and chiral field}\label{sec:decomp}

Contracting the propagator with the velocity and using $B^{ij}v_j=\hat k^i
(\hat{\mathbf k}\cdot\mathbf v)$, $A^{ij}v_j=v^i-\hat k^i(\hat{\mathbf k}\cdot
\mathbf v)$, and $C^{ij}v_j=i(\mathbf v\times\hat{\mathbf k})^i$, the induced
field $E^i=i\omega\Delta^{ij}Qv_j$ on the wake shell $\omega=\mathbf k\cdot
\mathbf v$ splits into the longitudinal, transverse, and chiral pieces of
Eqs.~(\ref{eq:EL})--(\ref{eq:Echi}) of the main text. The chiral field carries three exact
properties.

\textit{It sources no charge.} The induced density is $\varrho_{\rm ind}=i\mathbf
k\cdot\mathbf E$. For the chiral field $\mathbf k\cdot\mathbf E_\chi\propto\mathbf
k\cdot(\mathbf v\times\mathbf k)=0$, and for the transverse field $\mathbf k\cdot
\mathbf E_T\propto(\mathbf k\cdot\mathbf v)-\omega=0$ on shell, so the entire
induced charge density comes from $\mathbf E_L$ alone, consistent with Eq.~\eqref{eq:rho_pos} of
the main text.

\textit{It is solenoidal and odd in $\mu_5$.} Its curl, $\mathbf k\times(\mathbf v
\times\mathbf k)=k^2\mathbf v-(\mathbf k\cdot\mathbf v)\mathbf k$, is nonzero, so
the field is rotational rather than gradient-like. Under $\mu_5\to-\mu_5$ only
$C_A$ changes sign, while $C_L$, $C_T$, and $C_A^2$ are even, so $\mathbf E_\chi
\to-\mathbf E_\chi$.

\textit{It does no work on the parton.} The power delivered is $\mathbf v\cdot
\mathbf E_\chi\propto\mathbf v\cdot(\mathbf v\times\mathbf k)=0$, the field-level
form of the cancellation that keeps the soft polarization loss even in
$\mu_5$~\cite{ghosh_jamal_kurian}.

\section{Magnetic field and azimuthal current}\label{sec:bj}

In temporal gauge $\mathbf B=(\mathbf k\times\mathbf E)/\omega=iQ(\mathbf k\times
\mathbf D)$ with $\mathbf D\equiv\Delta\mathbf v$. Since $\mathbf k\times\mathbf
E_L=0$, only the transverse and chiral parts contribute,
\ba
\mathbf B_{\rm sym}&=&iQ\,\frac{C_T}{C_T^2-C_A^2}\,(\mathbf k\times\mathbf v),\\
\mathbf B_\chi&=&Q\,\frac{C_A}{C_T^2-C_A^2}\,\frac{1}{k}
\big(k^2\mathbf v-\omega\mathbf k\big),
\ea
the first, $\mathbf B_{\rm sym}$, being the parity-even (symmetric) part, the magnetic partner of $\mathbf E_T$ and even in $\mu_5$, and the second, $\mathbf B_\chi$, being Eq.\eqref{eq:Echi} of the main text. It lies in the plane of $\mathbf v$
and $\mathbf k$, hence is poloidal, and is odd in $\mu_5$. The induced current
follows from $\mathbf j=i(\mathbf k\times\mathbf B+\omega\mathbf E)$, giving
$j^i=-Q[\omega k^i/C_L+(\omega^2-k^2)D^i]$. Only the chiral part $\mathbf D_\chi$
has a nonvanishing azimuthal projection, so the azimuthal current reduces to
Eq.~\eqref{eq:jphi} of the main text, odd in $\mu_5$ and vanishing at $\mu_5=0$.

\section{Reduction to comoving coordinates}\label{sec:coord}

\begin{figure}[t]
  \centering
  \includegraphics[width=0.82\columnwidth,height=0.50\columnwidth]{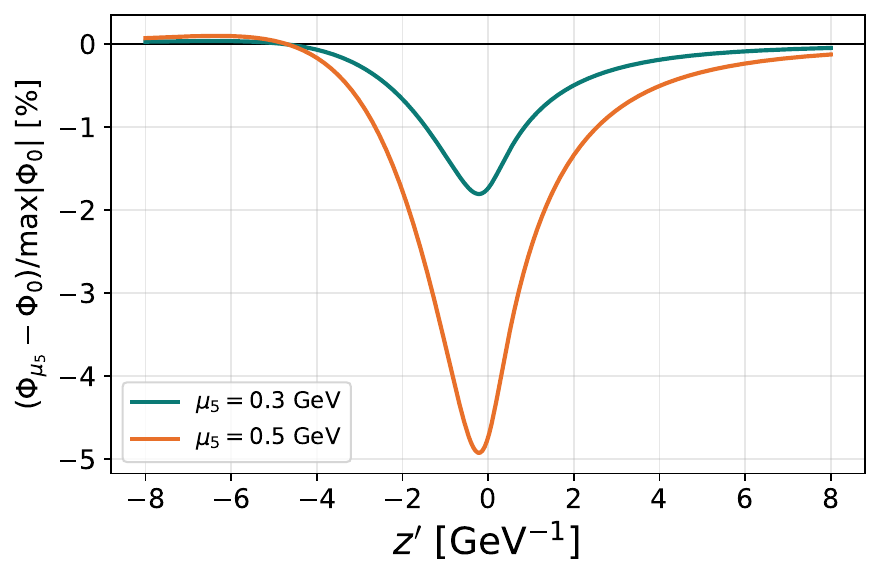}
  \caption{Fractional change of the induced screening potential,
  $(\Phi_{\mu_5}-\Phi_0)/\max|\Phi_0|$, along the symmetry axis as a function of
  $z'$, for $\mu_5=0.3$ and $0.5$~GeV. The change stays below one percent in magnitude, confirming that the screening sector responds to $\mu_5$ only through the Debye mass.}
  \label{fig:pfrac}
\end{figure}

With $\mathbf v=v\hat{\mathbf z}$, $\omega=vk_z$, and the comoving coordinate
$z'=z-vt$, axial symmetry reduces the inverse Fourier transform to a Hankel
transform in $k_\perp$ and a Fourier transform in $k_z$, through
\ba
\int_0^{2\pi}\! d\psi\,e^{ik_\perp\rho\cos\psi}&=&2\pi J_0(k_\perp\rho),\\
\int_0^{2\pi}\! d\psi\,e^{ik_\perp\rho\cos\psi}\cos\psi&=&2\pi i J_1(k_\perp\rho).
\ea
Applying this to the longitudinal field gives the screening potential and the
induced charge density,
\be
\Phi=\frac{Q}{(2\pi)^2}\mathrm{Re}\!\int\! dk_\perp k_\perp J_0\!\int\! dk_z\,
e^{ik_z z'}\Big[-\frac{\omega^2}{k^2 C_L}\Big],
\ee
with $\varrho_{\rm ind}=k^2\Phi$, Eq.~\eqref{eq:rho_pos} of the main text, depending on $\mu_5$
only through $C_L$ and hence only through $m_D$. The weak $\mu_5$ dependence of $\Phi$ is illustrated in Fig.~\ref{fig:pfrac}, where its fractional change stays below one percent over the range of $\mu_5$ studied. The transverse and chiral
projections give the azimuthal current, Eq.~\eqref{eq:jphi_pos}, and the meridional electric and
poloidal magnetic components, Eqs.~(\ref{eq:Ez})--(\ref{eq:Brho}), in which
\be
d_z=\frac{vk_z^2}{k^2 C_L}+\frac{C_T}{C_T^2-C_A^2}\frac{vk_\perp^2}{k^2},\quad
g_E=\frac{1}{k^2 C_L}-\frac{C_T}{k^2(C_T^2-C_A^2)}.
\ee

\begin{figure}[t]
  \centering
  \includegraphics[width=0.92\columnwidth,height=0.50\columnwidth]{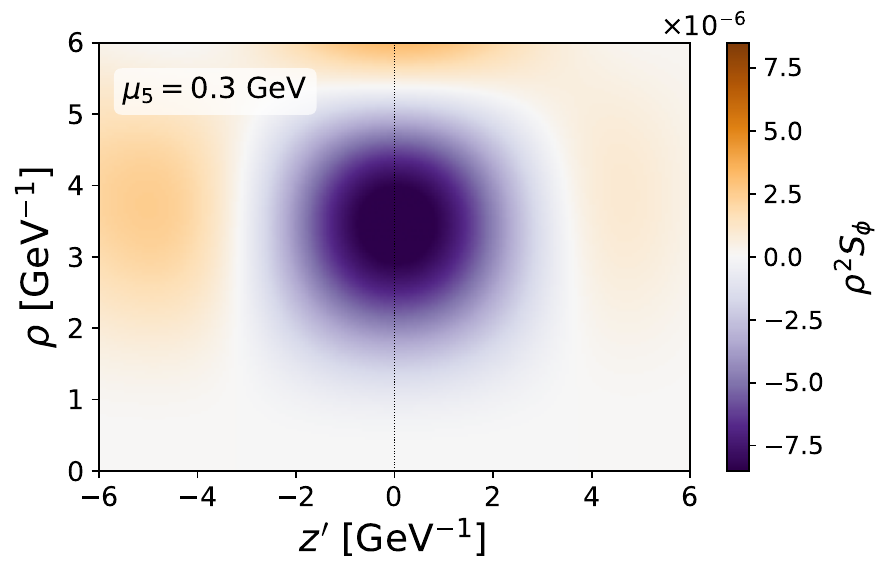}
  \caption{Angular-momentum density $\rho^2(E_z B_\rho-E_\rho B_z)$, the integrand
  of $L_z$ in Eq.~\eqref{eq:Lz} of the main text, in the comoving plane at $\mu_5=0.3$~GeV.
  The $\rho^2$ weight suppresses the contribution on the axis and concentrates it
  in a shell at intermediate $\rho$; the density is odd in $\mu_5$ and integrates
  to the quoted $L_z$. $v=0.55$, $k_{\max}=m_D$.}
  \label{fig:lzdens}
\end{figure}

These enter the field angular momentum, Eq.~\eqref{eq:Lz} of the main text. Because the
meridional $\mathbf E$ is even in $\mu_5$ and the poloidal $\mathbf B$ is odd, the
integrand $\rho^2(E_z B_\rho-E_\rho B_z)$ is odd in $\mu_5$ and vanishes at
$\mu_5=0$. Its spatial structure is shown in Fig.~\ref{fig:lzdens}: the $\rho^2$
weight suppresses the density on the axis and shifts the dominant contribution to
a shell at intermediate $\rho$, where the handed angular momentum is generated.


\begin{thebibliography}{99}
\bibitem{gale_review} C.~Gale, S.~Jeon, and B.~Schenke, Int. J. Mod. Phys. A \textbf{28}, 1340011 (2013).
\bibitem{heinz_snellings} U.~Heinz and R.~Snellings, Annu. Rev. Nucl. Part. Sci. \textbf{63}, 123 (2013).
\bibitem{romatschke} P.~Romatschke and U.~Romatschke, Phys. Rev. Lett. \textbf{99}, 172301 (2007).
\bibitem{bernhard} J.~E.~Bernhard, J.~S.~Moreland, and S.~A.~Bass, Nat. Phys. \textbf{15}, 1113 (2019).
\bibitem{jetscape} D.~Everett \textit{et al.} (JETSCAPE Collaboration), Phys. Rev. Lett. \textbf{126}, 242301 (2021).
\bibitem{mustafa2005} M.~G.~Mustafa, Phys. Rev. C \textbf{72}, 014905 (2005).
\bibitem{qin_prl} G.-Y.~Qin, J.~Ruppert, C.~Gale, S.~Jeon, G.~D.~Moore, and M.~G.~Mustafa, Phys. Rev. Lett. \textbf{100}, 072301 (2008).
\bibitem{cao} S.~Cao \textit{et al.}, Phys. Rev. C \textbf{99}, 054907 (2019).
\bibitem{sebastian_jamal} J.~Sebastian, M.~Y.~Jamal, and N.~Haque, Phys. Rev. D \textbf{107}, 054040 (2023).
\bibitem{ruppert_muller} J.~Ruppert and B.~M\"uller, Phys. Lett. B \textbf{618}, 123 (2005).
\bibitem{chakraborty_prd} P.~Chakraborty, M.~G.~Mustafa, and M.~H.~Thoma, Phys. Rev. D \textbf{74}, 094002 (2006).
\bibitem{chakraborty_prc} P.~Chakraborty, M.~G.~Mustafa, and M.~H.~Thoma, Phys. Rev. C \textbf{75}, 064908 (2007).
\bibitem{thoma_gyulassy} M.~H.~Thoma and M.~Gyulassy, Nucl. Phys. \textbf{B351}, 491 (1991).
\bibitem{wong} S.~K.~Wong, Nuovo Cimento A \textbf{65}, 689 (1970).
\bibitem{dumitru} A.~Dumitru, Y.~Nara, B.~Schenke, and M.~Strickland, Phys. Rev. C \textbf{78}, 024909 (2008).
\bibitem{carrington2015} M.~E.~Carrington, K.~Deja, and S.~Mr\'owczy\'nski, Phys. Rev. C \textbf{92}, 044914 (2015).
\bibitem{jamal_das_ruggieri} M.~Y.~Jamal, S.~K.~Das, and M.~Ruggieri, Phys. Rev. D \textbf{103}, 054030 (2021).
\bibitem{romatschke_strickland} P.~Romatschke and M.~Strickland, Phys. Rev. D \textbf{68}, 036004 (2003).
\bibitem{kharzeev_review} D.~E.~Kharzeev, J.~Liao, S.~A.~Voloshin, and G.~Wang, Prog. Part. Nucl. Phys. \textbf{88}, 1 (2016).
\bibitem{fukushima_kharzeev_warringa} K.~Fukushima, D.~E.~Kharzeev, and H.~J.~Warringa, Phys. Rev. D \textbf{78}, 074033 (2008).
\bibitem{kharzeev_son} D.~E.~Kharzeev and D.~T.~Son, Phys. Rev. Lett. \textbf{106}, 062301 (2011).
\bibitem{star_cme} J.~Adam \textit{et al.} (STAR Collaboration), Phys. Rev. Lett. \textbf{123}, 162301 (2019).
\bibitem{alice_cme} S.~Acharya \textit{et al.} (ALICE Collaboration), Phys. Rev. Lett. \textbf{125}, 022301 (2020).
\bibitem{stephanov_yin} M.~A.~Stephanov and Y.~Yin, Phys. Rev. Lett. \textbf{109}, 162001 (2012).
\bibitem{son_yamamoto} D.~T.~Son and N.~Yamamoto, Phys. Rev. Lett. \textbf{109}, 181602 (2012).
\bibitem{chen_ishii} J.-W.~Chen, T.~Ishii, S.~Pu, and N.~Yamamoto, Phys. Rev. D \textbf{93}, 125023 (2016).
\bibitem{akamatsu_yamamoto} Y.~Akamatsu and N.~Yamamoto, Phys. Rev. Lett. \textbf{111}, 052002 (2013).
\bibitem{carignano_manuel} S.~Carignano and C.~Manuel, Phys. Rev. D \textbf{103}, 116002 (2021).
\bibitem{ghosh_jamal_kurian} R.~Ghosh, M.~Y.~Jamal, and M.~Kurian, Phys. Rev. D \textbf{108}, 054035 (2023).
\bibitem{bellac} M.~Le Bellac, \textit{Thermal Field Theory} (Cambridge University Press, Cambridge, 2011).

\bibitem{star_polarization} L.~Adamczyk \textit{et al.} (STAR Collaboration), Nature (London) \textbf{548}, 62 (2017).
\bibitem{becattini_lisa} F.~Becattini and M.~A.~Lisa, Annu. Rev. Nucl. Part. Sci. \textbf{70}, 395 (2020).
\bibitem{liang_wang} Z.-T.~Liang and X.-N.~Wang, Phys. Rev. Lett. \textbf{94}, 102301 (2005); \textbf{96}, 039901(E) (2006).
\bibitem{alice_polarization} S.~Acharya \textit{et al.} (ALICE Collaboration), Phys. Rev. C \textbf{101}, 044611 (2020).
\bibitem{captain_vorticity} B.~Sahoo, C.~R.~Singh, and R.~Sahoo, arXiv:2604.13701.
\bibitem{miransky_shovkovy} V.~A.~Miransky and I.~A.~Shovkovy, Phys. Rep. \textbf{576}, 1 (2015).
\bibitem{carrington2017} M.~E.~Carrington, S.~Mr\'owczy\'nski, and B.~Schenke, Phys. Rev. C \textbf{95}, 024906 (2017).
\end{thebibliography}
\end{document}